# Chemical stability and superconductivity in Ag-sheathed CaKFe$_4$As$_4$ superconducting tapes

Zhe Cheng,[1,2] Chiheng Dong,[1,*] He Huang,[1,2] Shifa Liu,[1,2] Yanchang Zhu,[1,2] Dongliang Wang,[1] Vitalii Vlasko-Vlasov,[3] Ulrich Welp,[3] Wai–Kwong Kwok,[3] Yanwei Ma[1,2,*]

[1] *Key Laboratory of Applied Superconductivity, Institute of Electrical Engineering, Chinese Academy of Sciences, Beijing 100190, China*
[2] *University of Chinese Academy of Science, Beijing, 100049, China*
[3] *Argonne National Laboratory, 9700 South Cass Avenue, Argonne, Illinois 60439, USA*

\* E-mail address: dongch@mail.iee.ac.cn (C. Dong), ywma@mail.iee.ac.cn (Y. Ma)

**Abstract**

Ag-sheathed CaKFe$_4$As$_4$ superconducting tapes have been fabricated via the ex-situ powder-in-tube method. Thermal and X-ray diffraction analyses suggest that the CaKFe$_4$As$_4$ phase is unstable at high temperatures. It decomposes into the CaAgAs phase which reacts strongly with the silver sheath. We therefore sintered the tape at 500 °C and obtain a transport critical current density $J_c$(4.2 K, 0 T)~ $2.7 \times 10^4$ A/cm$^2$. The pinning potential derived from magnetoresistance measurements is one order of magnitude lower than that of the (Ba/Sr)$_{1-x}$K$_x$Fe$_2$As$_2$ tapes. Combining with the scanning electron microscopy and magneto-optical imaging results, we suggest that bad connectivity between superconducting grains caused by the low sintering temperature is the main factor responsible for the low $J_c$. However, this system is still a promising candidate for superconducting wires and tapes if we further optimize the post-annealing process to achieve better grain connectivity.

**Introduction**

The iron-based superconductors (IBSs) are of great interest from a basic point of view as well as in light of practical application [1]. In spite of the large varieties of IBSs with different crystal structures discovered so far, vortex pinning and anisotropy place fundamental restrictions on the current carrying ability. Furthermore, the high field performance of iron-based superconducting wires and tapes fabricated by the powder-in-tube (PIT) method [2] depends on other extrinsic factors, *i.e.* the purity of the precursors, the compatibility to the sheath material, the density and texture of the superconducting core, and particularly, the homogeneity of the superconducting phase. After ten years of research and design, the optimally doped (Ba/Sr)$_{1-x}$K$_x$Fe$_2$As$_2$ has emerged as the dominant material used in iron-based superconducting wires and tapes. On one hand, the quality of the (Ba/Sr)$_{1-x}$K$_x$Fe$_2$As$_2$ precursor has progressively improved due to mature synthesis processes [3]. On the other hand, plenty of technologies, including flat rolling [4,5], cold [6,7] and hot [8] uniaxial pressing, hot isotropic pressing [9–11], as well as double sheath architecture [12] have been applied in the manufacturing process to enhance the texture and density of the superconducting core. Based on these efforts, the critical current density $J_c$ of the (Ba/Sr)$_{1-x}$K$_x$Fe$_2$As$_2$ superconducting tapes at 4.2 K and 10 T has been gradually improved and now surpasses the level needed for practical applications [13]. Nevertheless, like other doped IBSs, the superconductivity of the BaK122 system is compromised by the non-uniformly distributed K atoms [14]. The inherent potassium clustering causes an inhomogeneous $J_c$ distribution due to its sensitivity to K content [15]. It deteriorates the overall current carrying capability of the iron-based superconducting wires and tapes, especially at long lengths. Consequently, finding a non-doped, stoichiometric iron-based superconductor becomes vitally important for practical applications.

The recently discovered CaKFe$_4$As$_4$ superconductor is prominent among stoichiometric IBSs owing to its high superconducting transition temperature of T$_c \sim$ 35 K [16] which is competitive with the optimally K doped FeAs122 system. In contrast to the (Ca$_{0.5}$Na$_{0.5}$)Fe$_2$As$_2$ solid solution with *I*4/*mmm* space group, CaKFe$_4$As$_4$ possesses a *P*4/*mmm* structure with the Ca and K layers alternately stacked between the Fe$_2$As$_2$ layers. Potassium doping on the alkaline earth element site unavoidably introduces substantial disorders whereas CaKFe$_4$As$_4$ is structurally ordered with Ca and K occupying different layers. In addition, there is no structural or magnetic phase transition [17], so one could consider this system as an optimally-doped ordered system. The upper critical field B$_{c2}$ of the CaKFe$_4$As$_4$ singe crystal is 71 T with the field parallel to the *c* axis and 92 T for the field perpendicular to the *c* axis [17]. The anisotropy parameter γ below 30 K is less than 2. The depairing current density $J_0$ at 0 K, estimated from $J_0 = \frac{\Phi_0}{3\sqrt{3}\mu_0\pi\xi\lambda^2}$, is 265 MA/cm$^2$, where $\Phi_0$ is the flux quantum, $\xi$=2.15 nm is the coherence length, λ=133 nm is the penetration depth [18]. This $J_0$-value is even larger than seen in other FeAs122 superconductors [19]. Further magnetic measurements on CaKFe$_4$As$_4$ single crystals indicate that the critical current density $J_c$ at 10 T is nearly 1 MA/cm$^2$ with the field parallel to the $c$ axis [20]. Moreover, it exhibits a more robust temperature dependence than the BaK122 counterpart, making it very promising to be used not only at liquid helium temperature, but also at intermediate temperatures accessible with cryocoolers.

In this paper, we synthesize the CaKFe$_4$As$_4$ precursor with a two-step method. The differential thermal analysis (DTA) on the precursor indicates a transition at 522 °C that is believed to be associated with the decomposition of the CaKFe$_4$As$_4$ phase. Therefore, we anneal the tape at 500 °C and achieve a $J_c$ of $2.7 \times 10^4$ A/cm$^2$ at 4.2 K and self-field. We demonstrate that bad connectivity between the superconducting grains due to low temperature sintering limits the overall critical current performance and indicate the need for a well-controlled post-annealing method to avoid the decomposition reaction in order to reach higher $J_c$ in this system.

**Experimental details**

In order to improve the homogeneity, we synthesized the precursor according to a two-step method [3]. We firstly prepared the intermediate compounds with nominal compositions CaAs and KAs at 700 °C and 400 °C, respectively. These intermediates were then mixed with Fe powders and As pieces according to the nominal composition Ca$_{1.15}$K$_{1.05}$Fe$_4$As$_4$. Excess CaAs and KAs were added to compensate for the loss of K and Ca during the heat treatment. The mixtures were ball milled for 10 hours in Ar atmosphere, sealed in a Nb tube afterwards and directly inserted into a preheated furnace at 900 °C. After 30 hours of sintering, the Nb tube was quenched to room temperature [16]. Considering the strong reaction between Cu/Fe and iron-pnictides, we chose Ag as the outer sheath material. After grinding the CaKFe$_4$As$_4$ precursor into powders, we loaded them into a silver tube with outer and inner diameters of 8 mm and 5 mm, respectively. The tube was swaged, drawn and rolled into a 0.4 mm thick tape. We cut the tape into short samples and finally annealed them at temperatures between 500 °C and 800 °C.

The X-ray diffraction (XRD) patterns of the precursor and the tapes were performed on a Bruker D8 Advance X-ray diffractometer. We analyzed the diffraction patterns by Rietveld refinement. The thermogravimetric (TG) and differential thermal analyses (DTA) were performed on a Synchronous thermal analyzer (Netzsch; STA 499 F3) at a heating rate of 20 °C/min. The average composition of the superconducting core is determined by an electron probe micro-analyzer (EPMA). The transport critical current $I_c$ was measured at 4.2 K via a four-probe method with a criterion of 1 µV/cm at the High Magnetic Field Laboratory (CHMFL, Hefei). We measured the resistivity of the superconducting core on a Physical Property Measurement System (PPMS). The microstructure of the tape was analyzed by scanning electron microscopy (SEM, Zeiss SIGMA). The magneto-optical image (MOI) of the superconducting core was obtained using an optical cryostat (Montana Instruments) at Argonne National Laboratory. We polished the superconducting core after tearing off the silver sheath on that side. A Bi-substituted iron-garnet indicator film was placed in direct contact with the superconducting core as magnetic field sensor [21]. The magnetic field generated from a homemade cooper coil is applied with the direction perpendicular to the tape surface.

**Results and discussion**

Fig.1 shows the X-ray diffraction pattern of the $CaKFe_4As_4$ precursor. We find that most of the diffraction peaks can be well indexed with the *P4/mmm* space group. There are additional reflections evidencing the presence of a small amount of residual Fe with *Im-3m* structure. However, we find no trace of $KFe_2As_2$ that is usually seen in the '1144' polycrystals. We thus perform Rietveld refinement with two phases. The refinement is successful as indicated by the reliability factors: $R_p$=5 %, $R_{wp}$=6.59 %. The fitted lattice parameters are *a*= 3.861 nm and *b*= 12.830 nm. The calculated fractions of the $CaKFe_4As_4$ phase and the Fe phase are 97.2 % and 2.8 %, respectively. The EPMA measurements on 20 points of the superconducting core indicate that the average composition is $Ca_{0.97}K_{1.02}Fe_4As_4$, close to the nominal composition.

So far, scientists have found various stable 1144 systems with chemical formula *AeA*$Fe_4As_4$, where *Ae* and *A* are the alkaline-earth and alkaline elements, respectively. The large difference between the radii of the *Ae* and the *A* ions is necessary for the formation of the 1144 structure [16]. Song *et al.* further utilized density functional theory to study the stability of the 1144 system [22] and found that the *P4/mmm* structure is sensitive to temperature and observe that $CaKFe_4As_4$ is only stable below a critical temperature of

T'~780 K (507 °C). At higher temperatures, the 122 phase with *I4/mmm* structure will dominate. In order to obtain a comprehensive understanding of the phase transition during the post-annealing process and evaluate a precise value of T', we perform thermal gravity (TG) and differential thermal analysis (DTA) on the precursor. Fig.2 shows the TG and DTA curves of the precursor under an argon atmosphere. The loss of the mass due to the evaporation of potassium is not as evident until the temperature is close to the melting temperature $T_m$~950 °C at which point there is a sharp endothermic peak. At intermediate temperature, there is a sharp exothermic peak at 522 °C, which is consistent with the theoretical prediction of the decomposition of the 1144 phase.

In order to verify this prediction, we annealed the CaKFe$_4$As$_4$ tapes in vacuum at different temperatures. As shown in Fig.3, all the diffraction peaks of the tape annealed at 500 °C for 0.5 hour can be well indexed with the *P4/mmm* space group. There is no evidence for secondary phase. One can see a great enhancement of the intensities of the (00*l*) peaks, suggesting strong *c*-axis texture. In order to quantitatively evaluate the *c*-axis texture, we utilize the Lotgering method to calculate $F=(\rho-\rho_0)/(1-\rho_0)$ [23], where $\rho = \sum I(00l)/\sum I(hkl)$, $\rho_0 = \sum I_0(00l)/\sum I_0(hkl)$. I and I$_0$ are the intensities of every peak for the textured and randomly oriented samples, respectively. The calculated *F* value for the Ca1144 tapes annealed at 500 °C is 0.565, which is intermediate between the Sr$_{0.6}$K$_{0.4}$Fe$_2$As$_2$ superconducting tape [24] (*F*~0.476) and the Ba$_{0.6}$K$_{0.4}$Fe$_2$As$_2$ superconducting tape [25] (*F*~0.590). The large F value of the CaKFe$_4$As$_4$ tape indicates that the cold working process including drawing and rolling can effectively align the grains with the *c*-axis perpendicular to the tape surface. When the post-annealing temperature increases up to 700 °C, obvious minor phases appear in the diffraction pattern, namely the KFe$_2$A$_2$ and the CaAgAs phases, as shown in Fig. 3. Moreover, sintering at 800 °C largely suppresses the intensity of the CaKFe$_4$As$_4$ diffraction peaks, making KFe$_2$As$_2$ the predominant phase. Combining with the theoretical calculation and the DTA results, we suggest that there is a transition temperature between 500 °C and 700 °C, at which CaKFe$_4$As$_4$ decomposes into CaFe$_2$As$_2$ and KFe$_2$As$_2$. Further continuous sintering above the transition temperature induces strong reaction between the Ag sheath and the CaFe$_2$As$_2$ at the interface. KFe$_2$As$_2$ is a superconductor with $T_c$~3.5 K [26], while CaAgAs is a topological insulator [27]. These phases deteriorate the superconductivity of CaK1144 at

4.2 K and decrease the overall supercurrent. Therefore, different from the K-doped FeAs122 tapes which have a proper sintering temperature widow ranging from 700 °C to 900 °C, we suggest that the CaKFe$_4$As$_4$ tapes can only be post-annealed below the critical temperature.

We measure the critical current of the CaKFe$_4$As$_4$ tapes annealed at different temperatures. As expected, all the tapes sintered at 700 °C and 800 °C shows no trace of supercurrents at 4.2 K. In contrast, we do observe a supercurrent in the tape post-annealed at 500 °C, as shown in Fig.4. At 0 T, the transport $J_c$ is 2.7 × 10$^4$ A/cm$^2$ (critical current $I_c$~129 A). With increasing field, the critical current density exhibits a strong field dependence and decreases to a value at 10 T that is nearly one-tenth of that at self-field. Furthermore, there is a clear hysteresis, indicating a weak-link effect in this sample. The transport $J_c$ at 4.2 K and 10 T is 2.2 × 10$^3$ A/cm$^2$. This value is one order of magnitude larger than the 1111 [28] and Co-doped 122 rolled tapes [29], and much smaller than that of the K-doped 122 rolled tapes [3].

The main panel of Fig. 5(a) shows the temperature dependence of normalized resistance at 0 T. The silver sheath is peeled off before the measurement. With decreasing temperature, the resistance exhibits a metallic like behavior and suddenly drops at $T_c$~35 K. The residual resistance ratio is RRR=$\rho$(300k)/$\rho$(35K) =4.65, close to that of the hot pressed Sr$_{0.6}$K$_{0.4}$Fe$_2$As$_2$ tapes [30]. However, the superconducting transition width $\Delta T_c$~5 K is rather large, which may be caused by bad connectivity. As shown in the insets of Fig.5(a), the $T_c^{onset}$ determined from the 90% criterion decreases very little with increasing field, while the $T_c^{zero}$ defined with the 10% criterion is largely suppressed by the field, resulting in an obvious tail feature near $T_c^{zero}$. Fig.5(b) shows the B-T phase diagram. One can see the very steep temperature dependence of B$_{c2}$ typically found in IBSs [31]. The slopes of the upper critical field, $-dB_{c2}/dT$, are 6.0 T/K and 9.7 T/K for the field perpendicular and parallel to the tape, respectively. These values are larger than that of the CaKFe$_4$As$_4$ single crystal, but close to the value of the powder sample [20]. The B$_{c2}$ at 0 K determined with the Werthamer-Helfand-Hohenberg (WHH) formula is 242 T and 110 T for the field parallel and perpendicular to the tape surface, respectively. The anisotropy parameter of B$_{c2}$ for fields applied perpendicular and parallel to the tape, $\gamma$= B$_{c2}^{//}$/ B$_{c2}^{\perp}$, is 2.32 near T$_c$ and decreases to 1.86 at 33.4 K, as shown in the inset of Fig.5(b).

The tail structure in the magnetoresistance curves may imply a possible thermal activated flux flow (TAFF) behavior. According to the TAFF model, the resistivity can be expressed as $\rho(T, H)=2\rho_c U\exp(-U/T)/T$, where U is the thermally activated energy or pinning potential. Assuming that $\rho_{0f} =2\rho_c U/T$ is a temperature independent constant, the thermal activation energy is $U=U_0(1-T/T_c)$, and we can obtain the Arrhenius relation:

$$\ln\rho(T,H)=\ln\rho_{0f}-U_0(H)/T+U_0(H)/T_c=\ln\rho_0-U_0(H)/T,$$

where $\ln\rho_0=\ln\rho_{0f}+U_0(H)/T_c$. The pinning potential can be evaluated from the slope, $U_0(H)=-d\ln\rho/d(1/T)$. We depict the field dependences of $U_0$ for the two field directions in Fig.5(c). The pinning potential at 0.5 T with the field perpendicular to the tape is 650 K. It is much smaller than the K-doped 122 single crystal and tape [13,32], implying weak thermal barrier against vortex motion. The pinning potential shows a weak power-law field dependence, namely $U_0 \propto B^{-n}$, with $n = 0.12$ and 0.14 for field perpendicular and parallel to the tape surface, respectively, manifesting a small anisotropy.

Fig.6(a) presents the optical image of the transverse section of the $CaKFe_4As_4$ tape showing a typical superconducting core with a saddle shape which arises due to the stronger deformation in the middle as compared to the two edges. Fig.6(b) shows a SEM image of the longitudinal section of the superconducting core. As a result of the low temperature sintering, the average size of the grains is as small as 2 μm. There are clearly observable voids and pores, limiting the actual current path along the tape. Fig.6(c) shows a MO image of the $CaKFe_4As_4$ superconducting tape. Magneto-optical images reveal the normal component of the magnetic induction, that is, the vortex density, at the imaged sample surface with a bright contrast representing high density. Note that in the current implementation this imaging technique does not distinguish between positive and negative vortices. The sample is cut into a rectangular shape with dimensions $4.08 \times 3.20 \times 0.18$ mm$^3$. We firstly cool the sample to 5 K in zero field. Then we increase the field to 82.7 mT and decrease it to zero to prepare the sample in the remanent state. We can see two bright areas corresponding to the thickest sections of the superconducting core which trap a high vortex density. The transverse feathering feature may be due to residual damage and cracking arising in the rolling process. The bright contrast along the center-line of the sample represents negative vortices which are generated by the return field from the vortices trapped in the thicker sample sections on either side.

**Conclusions**

In summary, we have fabricated Ag-sheathed CaKFe$_4$As$_4$ tapes by ex-situ PIT method. We find that CaKFe$_4$As$_4$ is unstable at high temperature. It tends to decompose into CaFe$_2$As$_2$, which strongly reacts with the silver sheath during the annealing process. Therefore, we sintered the tape at 500°C and obtained a self-field $J_c$ of $2.7 \times 10^4$ A/cm$^2$ at 4.2 K and self-field. The $J_c$ exhibits a strong field dependence and decreases to $2.2 \times 10^3$ A/cm$^2$ at 10 T. Magneto-transport and magneto-optical imaging reveal weak-link effects in the tape. The pinning potential derived from the magnetoresistance measurement is one order of magnitude smaller than that of the Ba$_{1-x}$K$_x$Fe$_2$As$_2$ tapes. Coupled with the SEM image, we conclude that bad connectivity between grains due to low temperature sintering is the main factor limiting the critical current at high fields. Considering the inherent high critical current density in the CaKFe$_4$As$_4$ single crystal, the polycrystalline wires and tapes are quite promising for practical application if a proper post-annealing process is applied to enhance the grain connectivity.

**Acknowledgements**

The authors would like to thank Dr. Fang Liu and Prof. Huajun Liu in High Magnetic Field Laboratory of the Chinese Academy of Sciences. We also thank Chen Li for help and useful suggestion. This work is partly supported by the National Natural Science Foundation of China (Grant No. 51320105015), the Beijing Municipal Science and Technology Commission (Grant No. Z171100002017006), the Bureau of Frontier Sciences and Education, Chinese Academy of Sciences (QYZDJ-SSW-JSC026), the Bureau of International Cooperation, Chinese Academy of Sciences (182111KYSB20160014) and Strategic Priority Research Program of Chinese Academy of Sciences (Grant No. XDB25000000). Magneto-optical work was carried out at Argonne National Laboratory and supported by the U.S. Department of Energy, Office of Science, Materials Sciences and Engineering Division.

045501

**Fig.1.** Powder XRD pattern and Rietveld refinement of the CaKFe$_4$As$_4$ precursor.

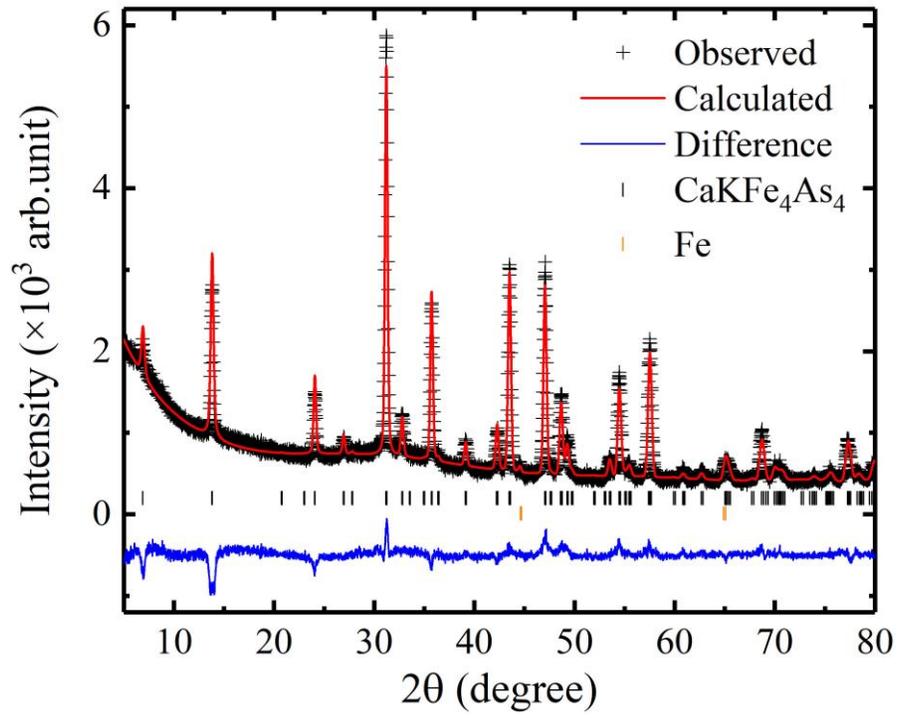

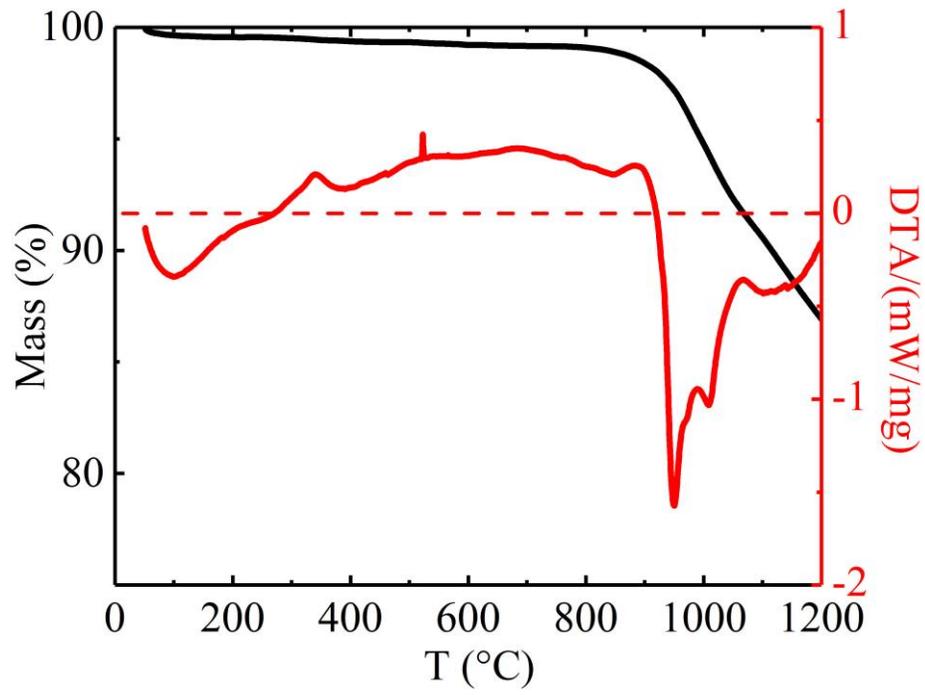

**Fig.2.** TG and DTA curves of the CaKFe$_4$As$_4$ precursor.

**Fig.3.** XRD patterns of the CaKFe$_4$As$_4$ superconducting tapes sintered at 500 °C, 700 °C and 800 °C.

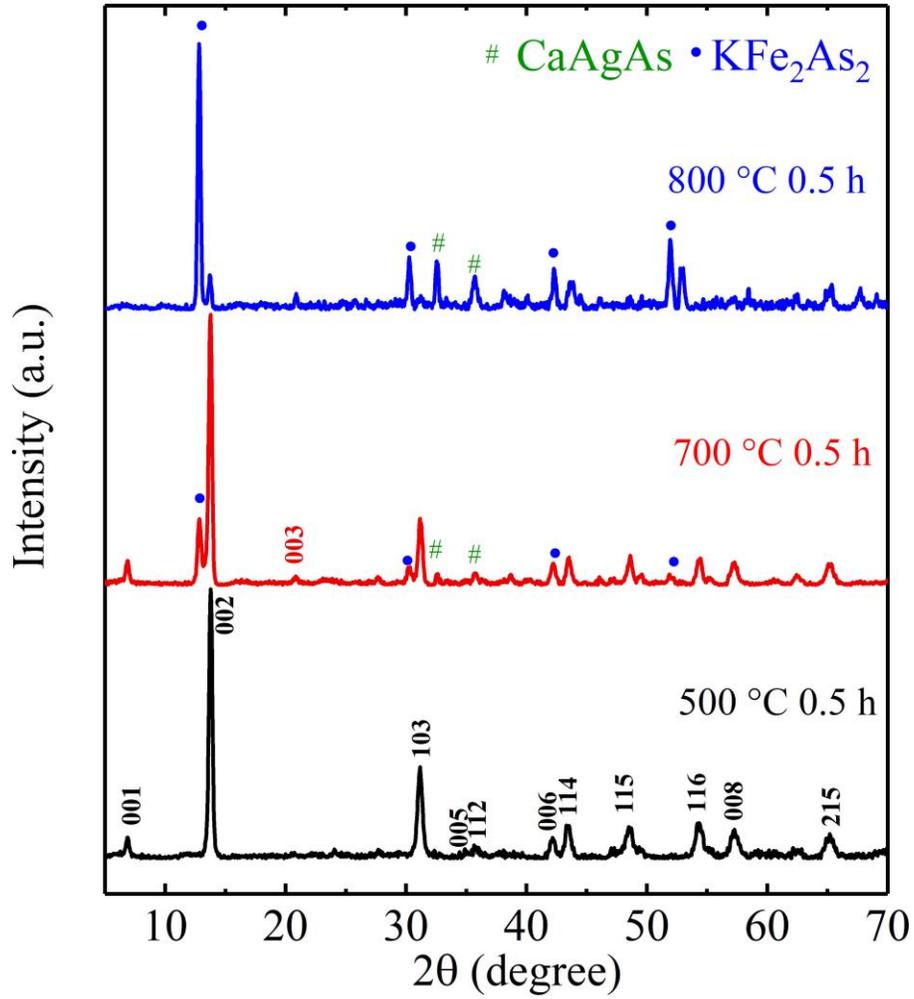

**Fig.4.** Magnetic field dependence of the transport $J_c$ at 4.2 K with increasing and decreasing field for the CaKFe$_4$As$_4$ tapes sintered at 500 °C for 0.5 h.

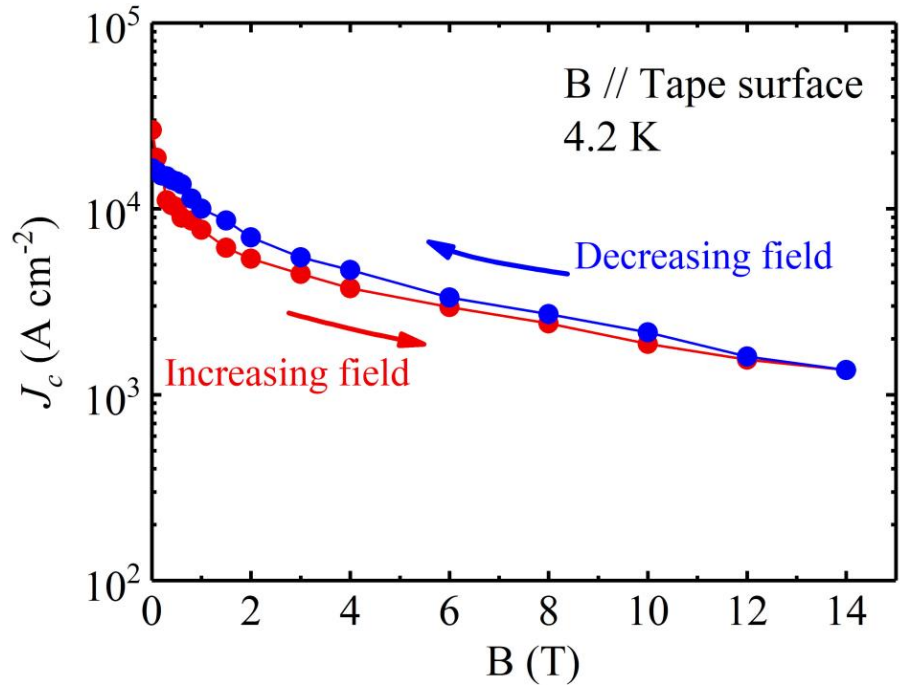

**Fig.5.** (a) Temperature dependence of the normalized resistance for the CaKFe$_4$As$_4$ tapes. Inset shows the magnetoresistance near T$_c$ with $H\perp$tape and $H \parallel$ tape. (b) B-T phase diagram, the inset is the anisotropy parameter as a function of temperature. (c) Field dependences of pinning potential for $H\perp$tape and $H \parallel$ tape.

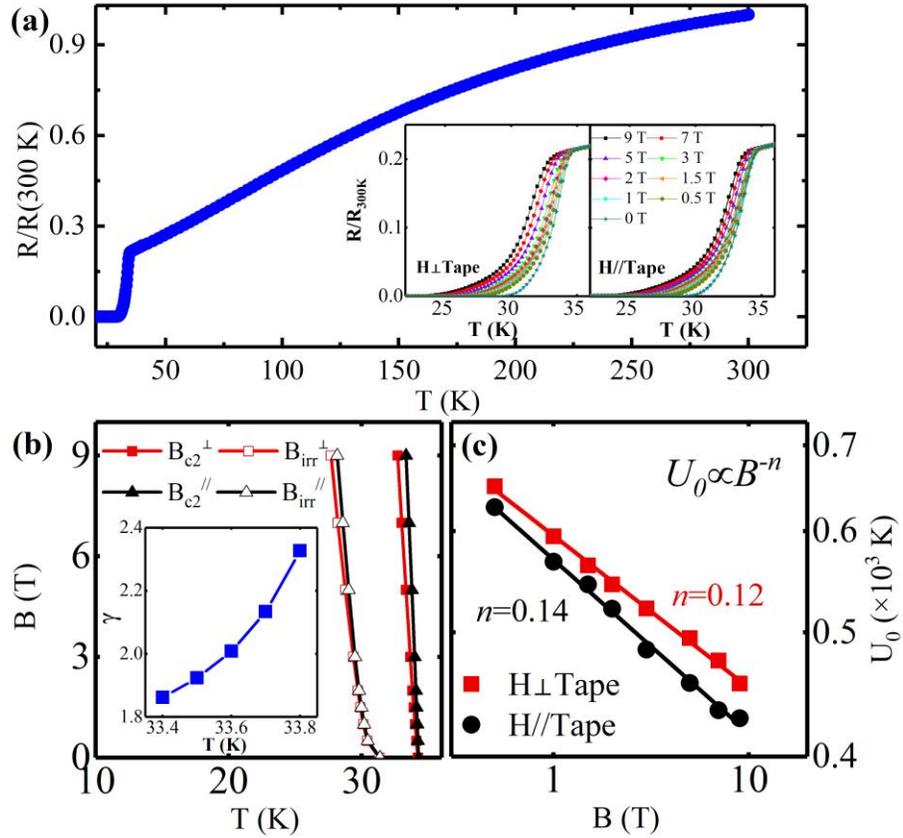

**Fig.6.** (a) Optical image of the transverse cross-section of the CaKFe$_4$As$_4$ tapes. (b) SEM image of the longitudinal section of the superconducting core. (c) MO image of the superconducting core. In this image the tape axis is vertical.

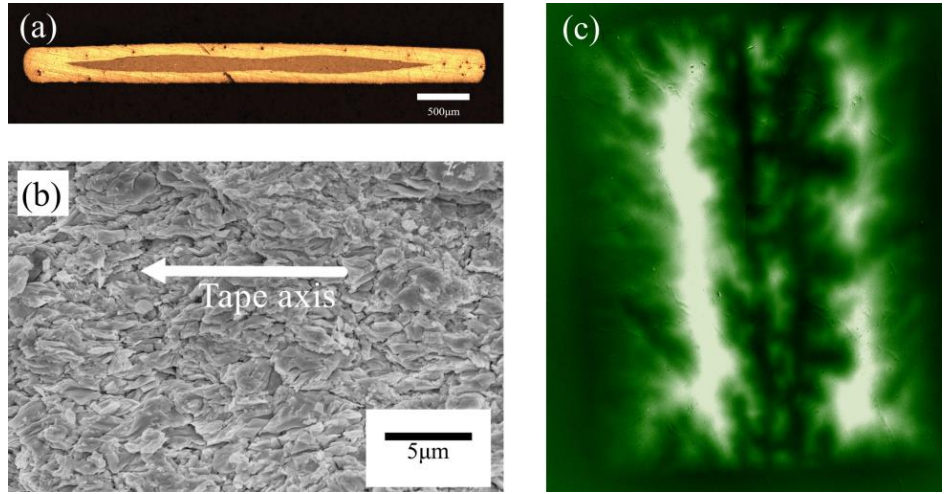